\documentstyle[12pt,a4,axodraw]{article}
\newcommand{\ee}{$e^-e^-$\ }
\newcommand{\ra}{\rightarrow}
\newcommand{\eennww}{$e^- e^- \rightarrow W^- W^- \nu_e \nu_e$}
\newcommand{\eewzen}{$e^- e^- \rightarrow W^- Z^0 e^- \nu_e$}
\newcommand{\eezzee}{$e^- e^- \rightarrow Z^0 Z^0 e^- e^-$}
\newcommand{\WW}{$W^-W^-\nu_e\nu_e$}
\newcommand{\WZ}{$W^-Z^0e^-\nu_e$}
\newcommand{\ZZ}{$Z^0Z^0e^-e^-$}
\newcommand{\bfig}{\begin{center}\begin{picture}}
\newcommand{\efig}{\end{picture}\end{center}}
\newcommand{\flin}[2]{\Line(#1)(#2)}

\newcommand{\zlin}[2]{\DashLine(#1)(#2){5}}
\newcommand{\glin}[3]{\Photon(#1)(#2){2}{#3}}

\newcommand{\sof}{\SetOffset}
\newcommand{\bea}{\begin{eqnarray}}
\newcommand{\eea}{\end{eqnarray}}

\begin{document}
\thispagestyle{empty}
\setcounter{page}{0}

\begin{flushright}
MPI-PhT/94-33\\
LMU-09/94\\
June 1994
\end{flushright}

\vspace*{15mm}

\begin{center}

{\Large \bf Search for new physics in \ee scattering\footnote{Talk presented
         by K. Ko\l odziej at the Zeuthen Workshop on Elementary Particle
         Theory, {\it `Physics at LEP200 and Beyond'}, Teupitz, Germany,
         10--15 April, 1994}\par}

\vspace{16mm}

{\large
F.~Cuypers$^{\rm a}$,
K.~Ko\l odziej$^{\rm b,}$\footnote{
        Work supported by the German Federal Ministry for Research
        and Technology under contract No.~05 6MU 93P}$^,$\footnote{
        On leave from the Institute of Physics, University of Silesia,
        ul.~Uniwersytecka 4, PL--40007 Katowice, Poland}
and R.~R\"uckl$^{\rm a,b,2}$} \\

\bigskip
{\em $^{\rm a}$Max-Planck-Institut f\"ur Physik, Werner-Heisenberg-Institut,\\
               F\"ohringer Ring 6, D--80805 M\"unchen, FRG}\\
\medskip
{\em $^{\rm b}$Sektion Physik der Universit\"at M\"unchen, \\
               Theresienstr.~37, D--80333 M\"unchen, FRG}\\

\end{center}

\vspace*{36mm}

\centerline{\bf Abstract}
\noindent
\normalsize
Considering the physics potential of an \ee collider in the TeV energy
range, we indicate a few interesting examples for exotic processes
and discuss the standard model backgrounds. Focussing on pair
production of weak gauge bosons, we report some illustrative
predictions.

\vfill
\newpage

\section{Introduction}

The prospects of experimentation at an $e^+e^-$ linear collider
facility are exciting, not the least because of the possibility
to collide in addition to $e^+$\ and $e^-$\ beams also $e^-$\
with $e^-$, $e^-$ with $\gamma$, and $\gamma$\ with $\gamma$\ beams.
Whereas much theoretical work
has already been dedicated to $e^+e^-$, $e\gamma$ and $\gamma\gamma$
physics \cite{Proc}, the \ee option has not yet been studied comprehensively.
One reason for this is the symmetry constraint in the standard
model which to lowest order allow only for
M\o ller scattering. However, it is just this fact which
makes \ee scattering very advantageous for new physics searches.

In this talk we indicate interesting nonstandard scenarios
which could be tested in \ee collisions. We also
discuss the main standard model backgrounds, except for the
rather harmless M\o ller scattering and Bremsstrahlung processes.
Moreover, we concentrate on the weak boson pair production processes
\eennww, \WZ\  and \ZZ\  for which we are able to report new results
\cite{CKKR,CKR}.

\section{New physics reactions}

A major virtue of \ee scattering as compared to the other
reactions to be achieved at a linear collider, is its
particular sensitivity to lepton-number violating processes.
For example, the existence of heavy Majorana neutrinos which
mix with the known neutrino species would increase the $W^-W^-$
pair production rate by inducing the reaction $e^-e^- \ra
W^-W^-$ \cite{Ng,Minkowski}. In particular, this process
is expected to occur in the framework of left-right symmetric
models, whose lagrangians naturally incorporate right-handed
neutrinos and Majorana mass terms, and in which the light
mass eigenstate $W^-$ is a mixing of $W_L^-$ and $W_R^-$.
The production of a heavy mass eigenstate $W^{'-}$ is also
an interesting possibility at higher energies \cite{Ng,Maalampi}.
Another example emerges in grand-unified
models which contain dileptons among the heavy gauge bosons.
These states give rise to electron-number violating transitions
of the kind $e^-e^- \ra \mu^-\mu^-$ \cite{Frampton}. The latter
process can also be mediated in left-right symmetric models
by doubly charged Higgs boson exchange and higher order
contributions from box diagrams involving Majorana neutrinos
and the charged gauge bosons $W^-_{L,R}$ \cite{Maalampi1}.
Furthermore, in supersymmetric models, lepton-number violating processes
can take place via the exchange of sneutrinos leading to
chargino pair production in
$e^-e^- \rightarrow \tilde{\chi}^- \tilde{\chi}^-$ \cite{COR}.

However, also lepton-number conserving processes can give insight
into new physics. For instance, the reaction $e^-e^- \ra e^-W^-\nu_e$
allows for tests of possible anomalous boson couplings which are
complementary to those in $e^+e^-$, $e\gamma$ and $\gamma\gamma$
collisions \cite{CC}. In addition, selectron pair production in
$e^-e^- \rightarrow \tilde{e}^- \tilde{e}^-$ proceeding by the
exchange of neutralinos, would provide a background-free
supersymmetric signal and a measurement of the neutralino mass
\cite{COR,KL}. Similarly, stronger constraints than in $e^+e^-$ can
be obtained for an additional $Z'$ boson \cite{CCL}.
In the following we will discuss some of the above reactions in more
detail.

\subsection{$e^-e^- \ra W^-W^-$}

Heavy Majorana neutrinos are naturally incorporated in
the framework of left-right symmetric gauge models \cite{Ng,Maalampi}
or $SO(10)$ GUTs \cite{Minkowski}. The production of $W^-$ pairs is then
mediated by $t$-channel exchange of Majorana neutrinos and $s$-channel
exchange of a doubly charged Higgs boson $\Delta^{--}$,
as depicted in Fig.~1. The latter mainly affects the high energy
behaviour of the cross sections due to interference with the
neutrino exchange amplitudes. The physical neutrino mass eigenstates
are mixtures of the known light neutrino flavours and the new heavy
states. Contributions from the exchange of light neutrino
eigenstates are suppressed by minute mixing angles of light
and heavy neutrinos (and/or of $W_L^-$ and $W_R^-$ in the case of
$e^-_Le^-_R$). These mixing angles are strongly constrained
by the nonobservation of neutrinoless double $\beta$-decay
\cite{Minkowski}.

Restricting oneself to the `low' energy region, $2m_W \ll \sqrt{s} \ll
m_{\nu_M}$, one can neglect the charged Higgs contribution.
Neglecting also $W_L^-$ and $W_R^-$ mixing,
the cross section for unpolarized electrons and heavy neutrino
exchange is estimated to be \cite{Minkowski}
\bea
\label{sigmaN}
    \sigma^N={{G_F^2s}\over {16\pi}}{s\over M_{\rm red}^2}
     \mid \eta_N \mid^2.
\eea
Here, $M_{\rm red}^{-1} = \sum_i M_i^{-1}$ denotes the inverse of the reduced
mass of heavy neutrinos which are assumed to have masses in the 1 TeV range,
and $\eta_N$ is a dimensionless neutrino mixing parameter.
For $\mid\eta_N\mid \leq (2-40)\;10^{-4}$, the maximal mixing compatible
with low energy data \cite{Minkowski}, one obtains the following
order of magnitude estimates:
\bea
      \sigma^N \leq \left\{ \begin{array}{ll}
       0.01-4\; {\rm fb} & \mbox{ at $\sqrt{s} = 0.5$ TeV} \\
       0.16-64\; {\rm fb} & \mbox{ at $\sqrt{s} = 1$ TeV.} \\
       \end{array} \right.
\eea

\subsection{$e^-e^- \ra \tilde{e}^- \tilde{e}^-$ or
                             $\tilde{\chi}^- \tilde{\chi}^-$}

In the supersymmetric extension of the standard model selectrons and
charginos are produced in pairs, as depicted in Fig.~2.

Selectron pair production proceeds via the $t$- and $u$-channel
exchange of neutralinos.
Note that this reaction also violates fermion-number conservation,
which comes as no surprise since the neutralinos are Majorana fermions.
The cross section for
$e^-e^- \ra \tilde{e}^- \tilde{e}^-$ depends very crucially
on the properties of the exchanged neutralinos \cite{LC18},
{\em i.e.} their masses and mixing.
In the minimal model, the masses $m_{\tilde\chi^0_i}$ of the four
neutralino states ($i=1,2,3,4$) and their couplings $g_{iL,R}$ to
electron-selectron pairs of left (L) and right (R) chiralities
are complicated functions of supersymmetry parameters \cite{LC3}.

Chargino pair production proceeds via the $t$- and $u$-channel
exchange of a sneutrino.
Since charginos only couple to left-handed leptons,
only the LL component of a given \ee initial state contributes.
Again, the masses $m_{\tilde\chi^-_i}$ of the charginos and
their couplings to selectrons $g_i$ are determined by the
supersymmetric parameters relevant also in the neutralino sector.

In Fig.~3, taken from Ref.~\cite{COR}, we show the energy dependence of
the selectron and chargino production cross sections for varying
selectron and sneutrino masses. For the ratio
of the Higgs vacuum expectation values we assume $v_2/v_1=\tan\beta=10$,
for the Higgs/higgsino mass parameter $\mu = -300$\ GeV and
for the mass parameter of the $SU(2)_L$
gaugino sector $M_2=300$ GeV. The corresponding
mass parameter $M_1$ of the $U(1)_Y$ gaugino sector is related to $M_2$
by evolving both from a common value $M_1 = M_2$ at the grand
unification scale, to the relevant low energy scale with the help
of renormalization group equations.

\section{Standard model processes}

As already mentioned in the Introduction, in the framework
of the standard model, the majority of events in \ee collisions arise
from M\o ller scattering accompanied by Bremsstrahlung:
\bea
e^-e^- \ra e^-e^-(\gamma ).
\eea
This background can be efficiently suppressed
by imposing acollinearity and/or acoplanarity cuts.

Next most important is the production and decay of
single gauge bosons in the reactions
\bea
e^-e^- &\ra&  W^-e^-\nu_e~,\\
e^-e^- &\ra&  Z^0e^-e^-~.
\eea
The cross sections for these processes
have been computed and discussed previously \cite{COR,CC}. They
are of the order of 1 pb.

Going still one power higher in the fine structure constant $\alpha$,
one encounters, among numerous other processes, pair production of
weak gauge bosons:
\bea
e^-e^- &\ra&  W^+W^-e^-e^-~,\\
e^-e^- &\ra&  W^-W^-\nu_e\nu_e~,\\
e^-e^- &\ra&  W^-Z^0e^-\nu_e~,\\
e^-e^- &\ra&  Z^0Z^0e^-e^-~.
\eea
Although these reactions do not take place very frequently
(at least at lower energies),
they may be a source of dangerous background to rare exotic
processes such as the examples contemplated in section 2.

In the following we will focus on the weak boson pair production
processes (7) to (9) \cite{CKKR,CKR}. Especially reaction (7) is
very important,
since it mimics the final state of $e^-e^- \ra W^-W^-$.
But also reactions (8) and (9), depending on the $W^-$ and $Z^0$
decay channels, may give rise to irreducible background.

The different topologies of Feynman diagrams contributing
to weak boson pair production are exemplified in Fig.~4.
In addition to double Bremsstrahlung diagrams
represented in Fig.~4a, there are diagrams involving the
triple gauge boson couplings as in Figs~4b, c, d and f,
gauge-Higgs boson couplings as in Figs~4e and g, and the
quartic gauge boson couplings as in Fig.~4h.
The actual diagrams can be easily constructed from these generic ones
by an appropriate particle assignment and by permutations
of the momenta of identical particles.
In the limit of zero electron mass,
one has 66, 88 and 86 diagrams in the \WW, \WZ\  and \ZZ\  channels,
respectively. In reaction (7), each of the two fermion lines is coupled
to a $W^-$ boson, so that only the $LL$ combination of initial
polarizations has a non-zero cross section. For reaction (8), also
the $LR$ combination of beam polarizations is relevant,
whereas the unpolarized cross section of reaction (9) receives
contributions from $LL$, $LR$\ and $RR$ beam polarizations.

When the traditional trace technique is employed,
not only does the number of terms become prohibitive,
but one also observes large gauge cancellations between them.
For these reasons,
we directly calculate
helicity amplitudes using two different methods and square them
numerically. In diagrams in which a virtual photon is coupled
to an on-shell electron line, the neglect of the electron
mass leads to collinear singularities. In these cases, regulating
terms proportional to $m_e^2$ are kept in the photon propagator.
Moreover, in order to obtain better numerical convergence of
the Monte Carlo integration, we introduce new variables which
smooth out the singular behaviour of the amplitudes in the collinear
regions. For further details of the calculation of the matrix elements
and the phase space integrations we refer to \cite{CKKR,CKR}.

In Fig.~5 we show numerical predictions\footnote{After this talk was
presented, we learned of two other recent calculations \cite{Boos} of
these processes, which confirm the results presented here.} for
the total unpolarized
cross sections of reactions (7) to (9) as a function of the \ee
centre-of-mass energy.
The following values are used for the electroweak parameters:
$m_Z = 91.19~{\rm GeV}$, $m_W = 80.3~{\rm GeV}$, $\sin^2\theta_W =
0.2244$, and $\alpha(m_Z^2) = 1/128.87$. Concerning $\alpha(Q^2)$,
there is some uncertainty related with the ambiguity of the choice
of the scale $Q^2$, a typical problem in $t$-channel processes.
For the Higgs boson mass we take $m_H = 100~{\rm GeV}$.
However, this choice has practically no influence on the numerical
results (except in reaction (9), where a Higgs boson heavier than
two $Z^0$ masses can be produced resonantly;
we do not contemplate this possibility here).
It is interesting to note the occurrence of subtle gauge
cancellations involving triple and quartic gauge boson couplings, which
reduce individual contributions from gauge invariant subsets
of Feynman diagrams by up to 3 orders of magnitude.

Assuming a realistic integrated luminosity of 10 fb$^{-1}$ per year,
we expect about 25 \WW, 100 \WZ, and 10 \ZZ\  events at $\sqrt{s} =500$
GeV. With increasing energy the yield of weak boson pairs becomes
much larger. This holds especially for reactions (7) and (8) the cross
sections of which rise to several tens to hundreds of fb above
1 TeV. Hence, comparing these rates with the cross sections for exotic
processes in Eq.~(2) and Fig.~3, one has to worry about the question
whether or not standard gauge boson pair production can obscure
interesting new physics.

Fortunately, a study of the
differential distributions for reactions (7--9) shows that this
background can be eliminated by imposing relatively simple kinematical
cuts. We illustrate this assertion for the process
$e_L^-e_R^- \ra W^-Z^0e^-\nu_e$ in Fig.~6 by displaying the distributions
of the transverse momentum and energy of the $W^-Z^0$ pair
at $\sqrt{s} = 0.5$ and 2 TeV.
We see that at $\sqrt{s} = 0.5$ TeV the energy distribution of the
$W^-Z^0$ pair assumes its maximum slightly above $0.5 \sqrt{s}$ whereas
the transverse momentum distribution of the gauge boson pair peaks
at about $0.1 \sqrt{s}$. At $\sqrt{s} = 2$ TeV, both distributions
are slightly softer. The peaking of the transverse
momentum distribution becomes more pronounced at $\sqrt{s} =2$ TeV
showing that the gauge bosons tend to be emitted more and more along
the beam axis as the energy increases. Another important observation
which can be exploited for discrimination
is the fact that, at both
centre-of-mass energies, the $W^-Z^0$ energy distribution dies out almost
completely for $E_{W^-}+E_{Z^0} > 0.8 \sqrt{s}$.
Thus, a simple cut imposed on the sum of gauge
boson energies should help to eliminate these processes as a background to
most of exotic reactions, in particular to  $e^-e^- \ra W^-W^-$,
where the gauge bosons would carry almost the whole centre-of-mass
energy.

These features are
independent of initial state polarization and hold for all
processes (7) to (9). Polarized cross sections and differential
distributions in other interesting variables are given in \cite{CKKR,CKR}.

\section{Conclusions}

The \ee option of a linear collider provides novel possibilities
to search for new physics. Especially lepton-number violating processes
can be probed with high sensitivity. The standard model backgrounds
mainly arise from single and pair production of weak gauge bosons.
The cross sections and differential distributions of these processes
can be evaluated with good accuracy. Hence, the expected numbers of
events can be estimated reliably. In addition, these backgrounds
can be substantially reduced by imposing relatively simple cuts
on angles and energies.

In the TeV energy range, the cross sections for
$e^-e^- \ra W^-e^-\nu_e~$, $e^-e^- \ra Z^0e^-e^-$, \eennww\  and \eewzen\
become large enough to allow for interesting studies of the gauge nature
of the standard model if luminosities of 10 fb$^{-1}$ are reached. Such
tests are particularly interesting in $W^-W^-$ and $W^-Z^0$ pair
production where triple and quartic gauge boson couplings contribute
at the same time.

\vfill
\eject

\vfill
\eject

\newpage

\begin{figure}[htb]
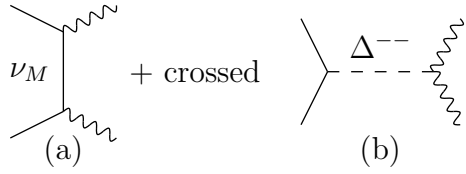

\bfig(150,60)
\sof(-10,10)
\Text(20,-5)[c]{(a)}
\flin{0,0}{20,10} \flin{20,10}{20,40} \flin{20,40}{0,50}
\glin{20,10}{40,0}{4} \glin{20,40}{40,50}{4}
\Text(15,25)[rc]{$\nu_M$}
\Text(45,25)[lc]{+ crossed}
\sof(100,10)
\Text(30,-5)[c]{(b)}
\flin{0,5}{10,25} \flin{10,25}{0,45}
\zlin{10,25}{50,25} \glin{50,25}{60,5}{4} \glin{50,25}{60,45}{4}
\Text(30,35)[c]{$\Delta^{--}$}
\efig
\caption[dummy]{Feynman diagrams for $e^-e^- \ra W^-W^-$; $\nu_M$ and
$\Delta^{--}$ denote a heavy neutrino mass eigenstate and a doubly
charged Higgs boson, respectively.}
\label{diag1}
\end{figure}

\vspace{1.cm}

\begin{figure}[htb]
\bfig(150,60)
\sof(-20,10)
\flin{0,0}{20,10} \flin{20,10}{20,40} \flin{20,40}{0,50}
\zlin{20,10}{40,0} \zlin{20,40}{40,50}
\Text(20,-5)[c]{(a)}
\Text(15,25)[rc]{$\tilde{\chi}_i^0$}
\Text(45,25)[lc]{+ crossed}
\Text(45,0)[lc]{$\tilde{e}^-$}
\Text(45,55)[lc]{$\tilde{e}^-$}
\sof(90,10)
\flin{0,0}{20,10} \zlin{20,10}{20,40} \flin{20,40}{0,50}
\flin{20,10}{40,0} \flin{20,40}{40,50}
\Text(20,-5)[c]{(b)}
\Text(15,25)[rc]{$\tilde{\nu}$}
\Text(45,25)[lc]{+ crossed}
\Text(45,0)[lc]{$\tilde{\chi}_j^-$}
\Text(45,55)[lc]{$\tilde{\chi}_i^-$}
\efig
\caption[dummy]{Feynman diagrams for the supersymmetric reactions
$e^-e^- \ra \tilde{e}^- \tilde{e}^-$ (a) and
$e^-e^- \ra \tilde{\chi}_i^- \tilde{\chi}_j^-$ (b).}
\label{diag2}
\end{figure}

\vfill
\eject

\newpage
\begin{figure}[t]
\centerline{
\begin{picture}(554,504)(0,0)
\put(0,0){\strut\epsffile{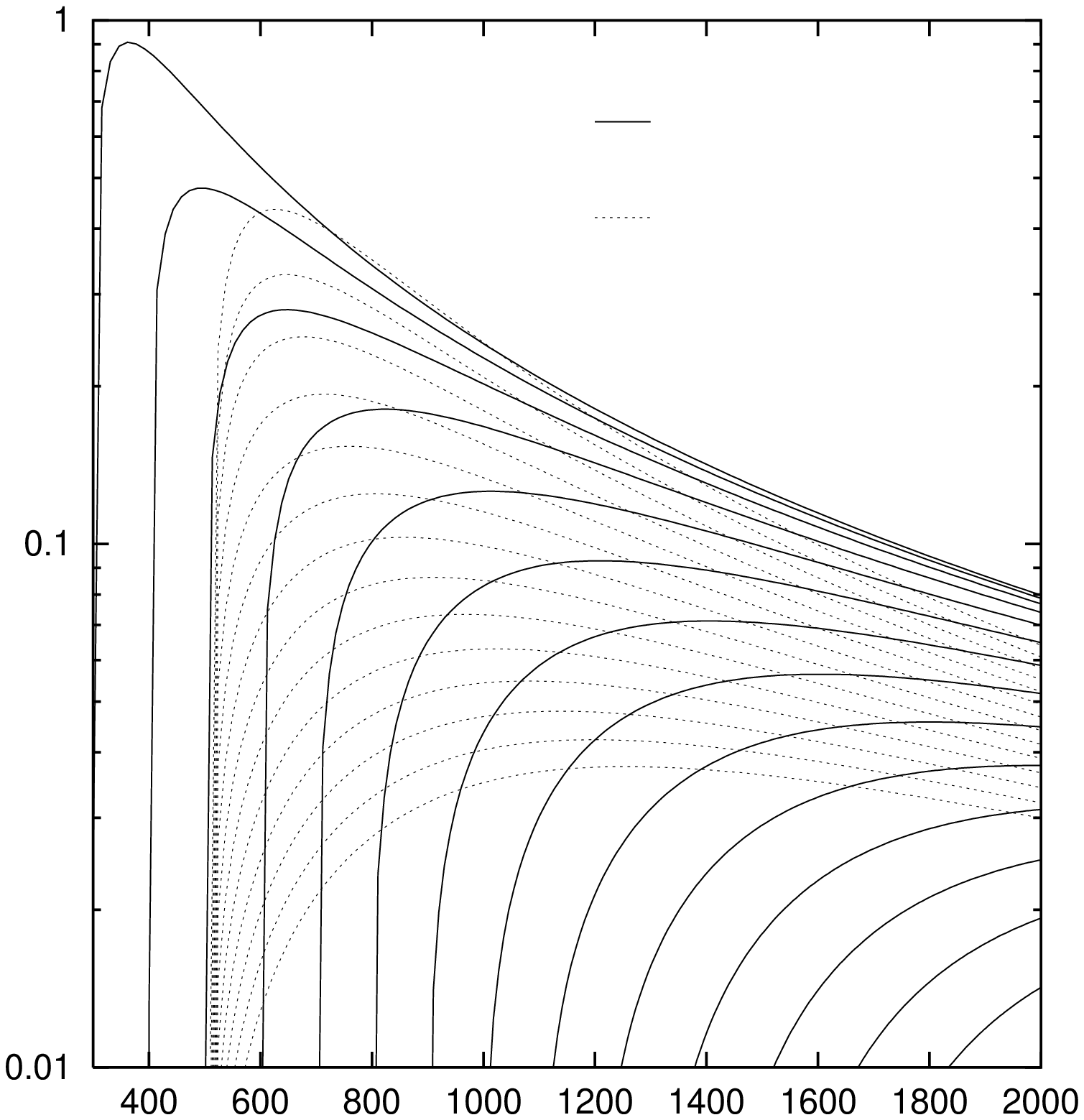}}
\put( 95.8,453.9){\makebox(0,0)[tr]{\Large$\sigma[{\rm pb}]$}}
\put(480.8, 19.1){\makebox(0,0)[tr]{\Large$\sqrt{s}[{\rm GeV}]$}}
\Text(345,430)[l]{\large $e^-e^-\to\tilde e^-\tilde e^-$}
\Text(345,390)[l]{\large $e^-e^-\to\tilde\chi_1^-\tilde\chi_1^-$}
\end{picture}
}
\caption[dummy]{Energy dependence of the unpolarized production cross
sections of $e^-e^-\ra\tilde e^-\tilde e^-$ (full curves)
and $e^-e^-\ra\tilde\chi_1^-\tilde\chi_1^-$ (dotted curves)
for $m_{\tilde e}=m_{\tilde \nu}=$150, 200, \ldots\ 800 GeV,
assuming $\tan\beta=10$, $\mu=-300$ GeV and $M_2=300$ GeV.
For this choice of parameters, $m_{\tilde\chi^-_1}=255$ GeV.
}
\label{eny}
\end{figure}

\begin{figure*}[htb]
\bfig(400,180)
\sof(0,110)
\flin{0,0}{60,0}
\flin{0,40}{60,40}
\glin{10,40}{30,60}{4}
\glin{20,40}{40,60}{4}
\glin{30,0}{30,40}{6}
\Text(30,-10)[c]{(a)}
\sof(80,110)
\flin{0,0}{60,0}
\flin{0,30}{60,30}
\glin{30,0}{30,30}{4}
\glin{15,30}{25,40}{3}
\glin{25,40}{30,60}{3}
\glin{25,40}{45,45}{3}
\Text(30,-10)[c]{(b)}
\sof(160,110)
\flin{0,0}{60,0}
\flin{0,40}{60,40}
\glin{30,0}{30,40}{6}
\glin{30,20}{50,20}{4}
\glin{15,40}{35,60}{4}
\Text(30,-10)[c]{(c)}
\sof(240,110)
\flin{0,0}{60,0}
\flin{0,40}{60,40}
\glin{30,0}{30,40}{6}
\glin{30,20}{45,20}{3}
\glin{45,20}{65,10}{4}
\glin{45,20}{65,30}{4}
\Text(30,-10)[c]{(d)}
\sof(320,110)
\flin{0,0}{60,0}
\flin{0,40}{60,40}
\glin{30,0}{30,40}{6}
\zlin{30,20}{45,20}
\glin{45,20}{65,10}{4}
\glin{45,20}{65,30}{4}
\Text(30,-10)[c]{(e)}
\sof(80,20)
\flin{0,0}{60,0}
\flin{0,60}{60,60}
\glin{30,0}{30,20}{3}
\glin{30,20}{30,40}{3}
\glin{30,40}{30,60}{3}
\glin{30,20}{60,20}{4}
\glin{30,40}{60,40}{4}
\Text(30,-10)[c]{(f)}
\sof(160,20)
\flin{0,0}{60,0}
\flin{0,60}{60,60}
\glin{30,0}{30,20}{4}
\zlin{30,20}{30,40}
\glin{30,40}{30,60}{4}
\glin{30,20}{60,20}{4}
\glin{30,40}{60,40}{4}
\Text(30,-10)[c]{(g)}
\sof(240,20)
\flin{0,0}{60,0}
\flin{0,60}{60,60}
\glin{30,0}{30,30}{4}
\glin{30,30}{30,60}{4}
\glin{30,30}{60,20}{4}
\glin{30,30}{60,40}{4}
\Text(30,-10)[c]{(h)}
\efig
\vspace*{-0.5cm}
\caption[dummy]{Typical topologies of the Feynman diagrams
                   of weak boson pair production reactions.
}
\label{diag3}
\end{figure*}

\vspace{-0.5cm}

\begin{figure}[t]
\centerline{
\begin{picture}(482,352)(0,0)
\put(0,30){\strut\epsffile{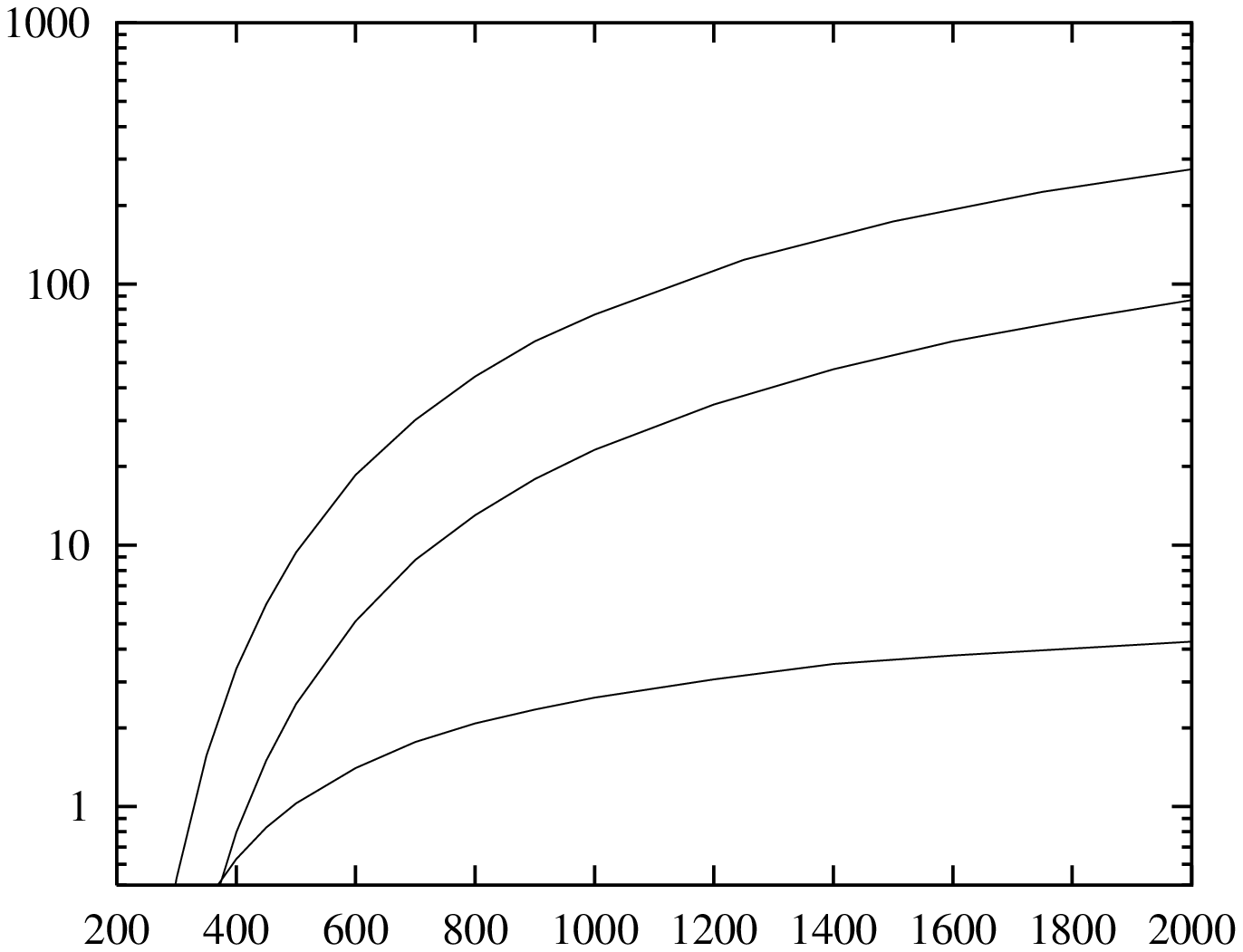}}
\put( 45.8,293.9){\makebox(0,0)[tr]{\Large $\sigma[{\rm fb}]$}}
\put(420.8, 12.){\makebox(0,0)[tr]{\Large $\sqrt{s}[{\rm GeV}]$}}
\Text(220,287)[l]{\large \eewzen}
\Text(220,180)[l]{\large \eennww}
\Text(220,99)[l]{\large \eezzee}
\end{picture}
}
\caption{Total cross sections for gauge boson pair production
	with unpolarized $e^-$ beams.}
\label{cstot}
\end{figure}

\vfill
\eject

\newpage
\begin{figure}[t]
\centerline{
\begin{picture}(482,650)(0,0)
\put(0,380){\strut\epsffile{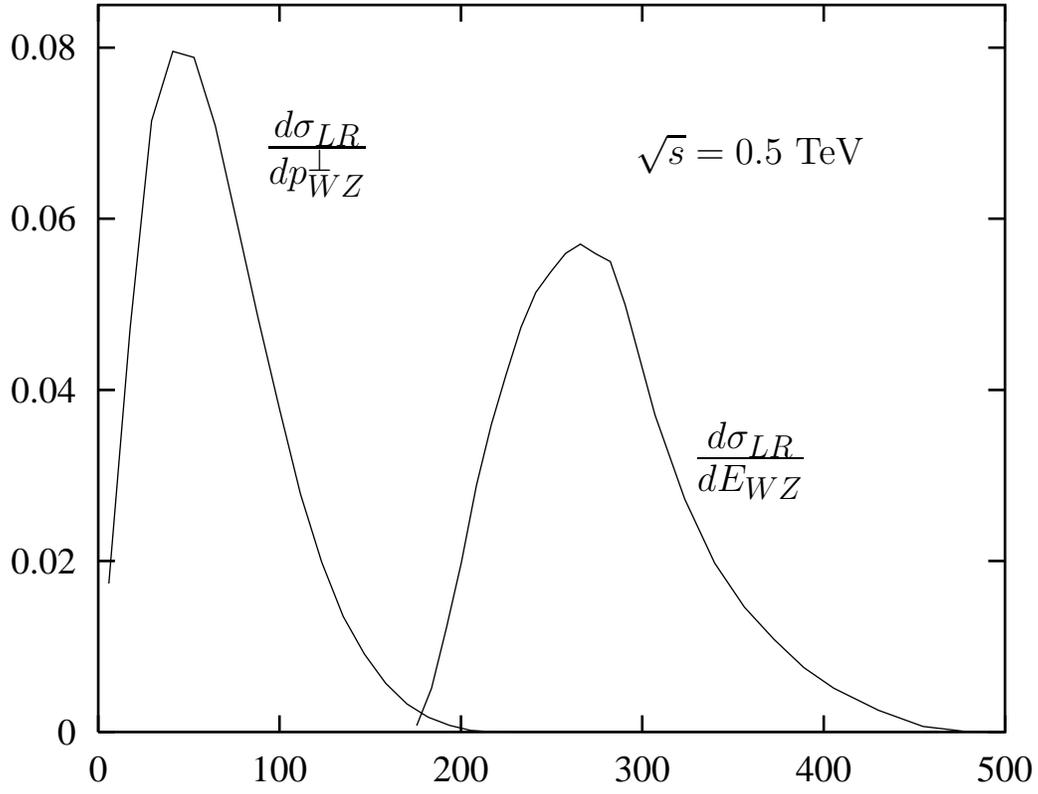}}
\put(0,40){\strut\epsffile{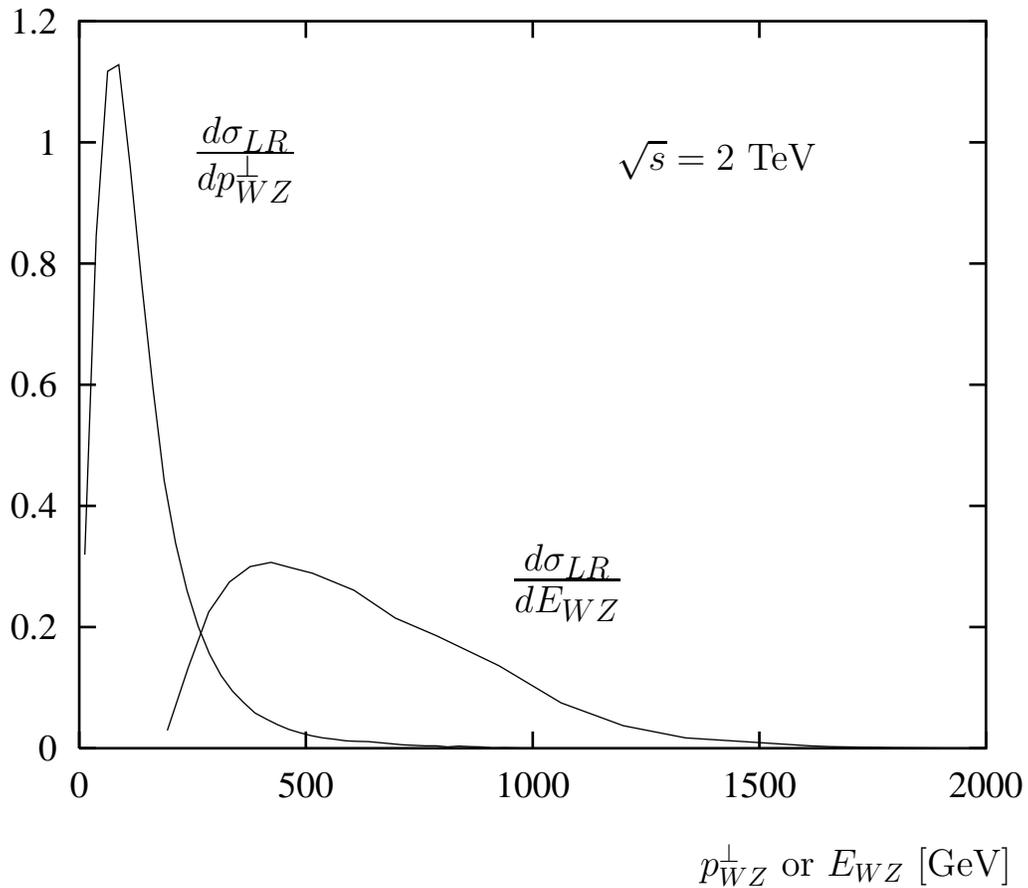}}
\put(130.8,620.0){\makebox(0,0)[l]{\LARGE ${d\sigma_{LR}
                                                    \over dp^\perp_{WZ}}$}}
\put(270.8,620.0){\makebox(0,0)[l]{\large $\sqrt{s} = 0.5$\ TeV}}
\put(293.0,503.9){\makebox(0,0)[l]{\LARGE ${d\sigma_{LR} \over dE_{WZ}}$}}
\put(110.8,283.9){\makebox(0,0)[l]{\LARGE ${d\sigma_{LR}
                                                    \over dp^\perp_{WZ}}$}}
\put(270.8,283.9){\makebox(0,0)[l]{\large $\sqrt{s} = 2$\ TeV}}
\put(230.9,123.9){\makebox(0,0)[l]{\LARGE ${d\sigma_{LR} \over dE_{WZ}}$}}
\put(420.,25.){\makebox(0,0)[tr]{\large $p^\perp_{WZ}$ or $E_{WZ}$ [GeV]}}
\end{picture}
}
\vspace{-15pt}
\caption{Transverse momentum and energy distributions of the
	$W^-Z^0$-pair in $e^-_Le^-_R \ra W^-Z^0e^-\nu_e$\ in
        fb/GeV.}
\label{wz}
\end{figure}

\end{document}